\documentclass[numsec,webpdf,modern,medium]{oup-authoring-template}

\onecolumn 

\graphicspath{{Fig/}}

\theoremstyle{thmstyleone}%
\theoremstyle{thmstyletwo}%
\theoremstyle{thmstylethree}%

\usepackage{amsmath}
\usepackage{graphicx}
\usepackage{url}
\usepackage{natbib}
\usepackage{listings}
\usepackage{bbm}
\usepackage{bm}
\usepackage{color}
\usepackage[mathscr]{euscript}
\newcommand{\cond}{\,\vert\,}
\newcommand{\doo}{\textrm{do}}
\newcommand{\moo}{\textrm{mo}}
\newcommand{\E}{\textrm{E}}

\usepackage{tikz}
\usetikzlibrary{arrows.meta, positioning}
\tikzset{mycircle/.style={circle, draw, minimum size=1cm}} 
\tikzset{myrectangle/.style={rectangle, draw, minimum size=1cm}}

\begin{document}

\journaltitle{Journal Title Here}
\DOI{DOI added during production}
\copyrightyear{YEAR}
\pubyear{YEAR}
\vol{XX}
\issue{x}
\access{Published: Date added during production}
\appnotes{Paper}

\firstpage{1}


\title[Short Article Title]{A context-specific causal model for estimating the effect of extended length of overnight stay on traveller's total expenditure}

\author[1,2,$\ast$]{Lauri Valkonen \ORCID{0000-0001-8332-4526}}
\author[2]{Juha Karvanen}

\address[1]{\orgdiv{HAMK Tech}, \orgname{Häme University of Applied Sciences}, \orgaddress{\street{Vankanlähde 9}, \postcode{13100}, \state{Hämeenlinna}, \country{Finland}}}
\address[2]{\orgdiv{Department of Mathematics and Statistics}, \orgname{University of Jyväskylä}, \orgaddress{\street{P.O. Box 35}, \postcode{40014}, \state{Jyväskylä}, \country{Finland}}}

\corresp[$\ast$]{Corresponding author. \href{email:email-id.com}{lauri.valkonen@hamk.fi}}


\abstract{Tourism significantly affects the economies of many countries.
Understanding the causal relationship between the length of overnight stay and traveller's expenditure is crucial for stakeholders to characterize spending profiles and to design marketing strategies. Causal mechanisms differ between personal and work-related travel because the decision-making processes have different drivers and constraints.
We apply context-specific independence relations to model causal mechanisms in contexts specified by trip purpose and identify the causal effect of the length of stay on expenditure.
Using the international visitor survey data on foreign travellers to Finland, we fit a hierarchical Bayesian model to estimate the posterior distribution of the counterfactual expenditure due to  extending the length of stay by one night. We also perform a Bayesian sensitivity analysis of the estimated causal effect with respect to omitted variable bias.} 

\keywords{Bayesian model, Causal Inference, Context-specific Independence, Expenditure, Tourism}





\maketitle


\section{Introduction}
Tourism comprises a major share of the global economy between countries, covering a wide variety of different industries from transportation to accommodation and catering services, as well as event management. In 2024, tourism represented approximately 10 \% of the gross domestic product (GDP) worldwide \citep{statista}. Among the various factors that influence economic impact and tourist behaviour, the length of stay is an essential determinant of the total expenditure of a traveller.

Extending the length of stay may contribute to an increase in overall expenditure by allowing travellers to consider additional consumable services and experiences. However, the relationship between the length of stay and the total expenditure may not be linear. For example, during a longer stay, a traveller may prefer cheaper alternatives for accommodation and activities.

Identifying the number overnight stays at which an additional night leads to the highest increase in the total expenditure enables stakeholders, such as businesses and destination management organizations, to tailor their efforts within the tourism market. In addition to the economic point of view, the relationship between the length of stay and sustainability objectives has gained increasing attention. Generally, promoting a longer length of stay over multiple shorter trips can help to decrease transport emissions (e.g. \cite{Gossling02042016}), but also potentially mitigate overcrowding in the destinations (e.g. \citet{oklevik2021determinants}).

The associations between the length of stay and traveller expenditure 
has been extensively studied \citep{BRIDA201328}. For example, \citet{FIEGER2021102974} reported a non-linear relationship between total expenditure and the length of stay, with total expenditure exhibiting a steeper grow for shorter-duration trips and then decelerating for longer stays. Furthermore, a saturation effect in the tourist expenditure has been suggested with the expenditure diminishing after a certain length of stay \citep[e.g.,][]{wang2018}.

The need for understanding causality in the context of travelling has been already recognized \citep{Mazanec01092007, BRATHWAITE20181}. Traditional methods for assessing causal questions in tourism typically rely on experimental frameworks, for which \citet{VIGLIA2020102858} use the categorization between laboratory experiments, field experiments, natural and quasi experiments, and discrete choice experiments, each having their own pros and cons. The previous experiment types are commonly employed in business-related questions, such as purchase intentions, consumption of services and supplies, as well as destination choice. However, inspecting the effect of the length of stay poses challenges for experimental research due to the limited control of the intervention variable. It is not possible to intervene directly on travellers' choice regarding the length of stay. Causal inference based on observational data \citep{pearl2009} is warranted given that a significant proportion of data generated within the tourism sector stems from passive observations. 

The purpose of the trip is a context that strongly impacts other choices the traveller makes related to the trip. For example, a traveller attending a conference will select the destination based on the location of the event. This makes the destination dictated by the work requirements and thus removes the causal pathway from personal preferences to the choice of the destination. On the other hand, individuals travelling for personal purposes usually make the decision about destination based on their personal preferences, which implies a causal path between these variables in this context.%

In this paper, we aim to estimate the effect of extended length of stay on traveller’s total expenditure using causal inference \citep{pearl2009}. The ultimate estimand of interest is the average counterfactual increase in total expenditures if a traveller had stayed one night longer than they actually did. %

We use a labelled directed acyclic graph (LDAG) \citep{pensar2015labeled} to form a context-specific causal model \citep{NEURIPS2019d88518ac} that allows us to take into account the difference between work-related trips and personal trips when identifying and estimating causal effects. Context-specific independence (CSI) relations \citep{csiboutulier} extend do-calculus \citep{pearl1995} and can enable the identification of causal effects that would otherwise remain unidentifiable \citep{NEURIPS2019d88518ac, mokhtarian2022causal}. Our study seems to be the first to apply CSI-relations in causal analysis of real data.%

We use the context-specific causal model to identify the counterfactual effect of the extended length of stay on the traveller’s total expenditure based on the survey data of foreign travellers visited in Finland. We fit a hierarchical Bayesian model to estimate the posterior distribution of this quantity. We also propose a Bayesian approach for sensitivity analysis for estimating the omitted variable bias regarding traveller’s annual income which has not been measured. We adopt ideas from \citet{cinelli2020making} and calculate the bias in the causal effect estimates based on the prior understanding on the relationships between the income and length of stay and between the income and total total expenditure in both contexts.%

The structure of the paper is the following. In Section \ref{CSImodel}, we outline the framework of the context-specific causal models.  Section \ref{data} describes the data set used in the analysis and the model for the data-generating process. In Section \ref{sec_identif} we leverage CSI-relations to derive a formula for the causal effect of extended length of stay. Subsequently, in Section \ref{estimation}, we fit a Bayesian model to estimate he total expenditure in the counterfactual case where the length of stay is extended by one night. Following this, in Section \ref{ovb} we present the sensitivity analysis to investigate the robustness of estimates to unobserved income. Finally, in Section \ref{discussion} we discuss the strengths and limitations of the proposed approach and the data used.

\section{Introduction to context-specific causal models} \label{CSImodel}

\begin{figure}[ht!!]
    \centering
    \begin{minipage}{0.45\textwidth}
\centering
\begin{tikzpicture}[>=Latex, fill=white] \label{exg1}

\node[draw, fill=white, mycircle] at (0,0) (M) {$M$};
\node[draw, fill=white, myrectangle] at (3,0) (U) {$U$};
\node[draw, fill=white, mycircle] at (0,-3) (X) {$X$};
\node[draw, fill=white, mycircle] at (3,-3) (Z) {$Z$};
\node[draw, fill=white, mycircle] at (1.5,-5) (Y) {$Y$};

\draw[->] (M) -- (X);
\draw[->] (M) -- (Z);
\draw[->] (X) -- (Y);
\draw[->] (X) -- (Z);
\draw[->] (Z) -- (Y);
\draw[->] (U) -- (X);
\draw[->] (U) -- (Z);

\end{tikzpicture}
    \end{minipage}
    \hspace{1cm}
\begin{minipage}{0.45\textwidth}
\centering
\begin{tikzpicture}[>=Latex, fill=white]

\node[draw, fill=white, mycircle] at (0,0) (M) {$M$};
\node[draw, fill=white, myrectangle] at (3,0) (U) {$U$};
\node[draw, fill=white, mycircle] at (0,-3) (X) {$X$};
\node[draw, fill=white, mycircle] at (3,-3) (Z) {$Z$};
\node[draw, fill=white, mycircle] at (1.5,-5) (Y) {$Y$};

\draw[->] (M) -- (X);
\draw[->] (M) -- (Z);
\draw[->] (X) -- (Y);
\draw[->] (Z) -- (Y);

\draw[dashed, ->] (U) -- node[midway, above, text opacity=1, fill=white] {$M=1$} (X);
\draw[dashed, ->] (U) -- node[midway, above, text opacity=1, fill=white] {$M=1$} (Z);
\draw[dashed, ->] (X) -- node[midway, above, text opacity=1, fill=white] {$M=0$} (Z);

\end{tikzpicture}
\end{minipage}
\caption{DAG (left) and LDAG (right) for illustrating the same data generating process, where additional context specific information resulting from node $M$ makes the query $P(Y \cond \doo(X))$ to be identifiable. In the graph nodes indicate the observed (circles) and unobserved variables (squares) and the edges represent the causal directions. Dashed edges with notation indicate the vanishing causal directions in the specific context. A circle node with an inner circle points out to the variable to be intervened.}
\label{fig:ex_graphs}
\end{figure}
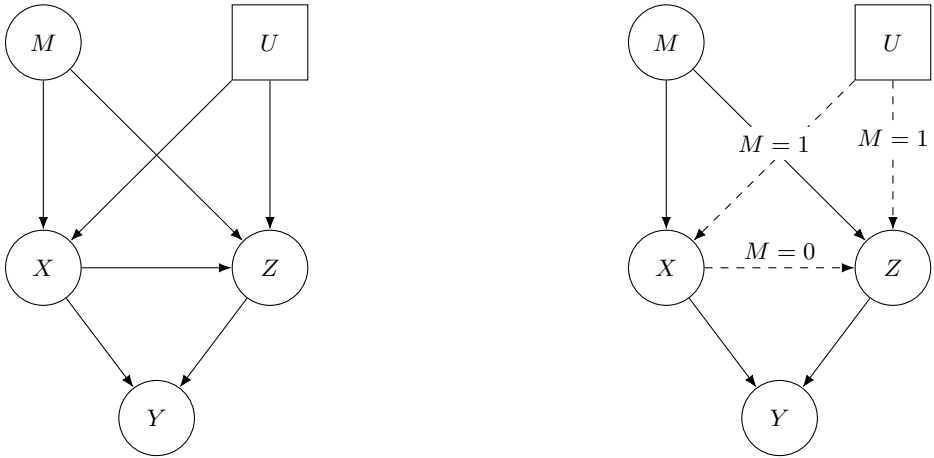

We start with a demonstration of the use of CSI-relations in the causal effect identification problem. Consider causal graphs presented in Figure~\ref{fig:ex_graphs}, containing observed variables $X$, $Y$, $Z$, $M$, and an unobserved variable $U$.  In the left causal graph the causal effect $P(Y \cond \doo(X))$ is non-identifiable because the unobserved confounder $U$ affects $X$ but also $Y$ through $Z$ which is a child of $X$. The situation changes in the right causal graph where CSI-relations are taken into account.

Let us then assume that we have context-specific information about the causal relationship between $X$, $Z$, and $U$, which is presented by the LDAG in the right panel of Figure \ref{fig:ex_graphs}. In the LDAG the dashed edges describe the causal directions that vanish in the specific context indicated by the context label. By applying the Do-search identification algorithm \citep{dosearchtikka2021,NEURIPS2019d88518ac}, we obtain the identifying functional

\begin{align}
\begin{split}
P(Y \cond \doo(X)) = &\sum_{Z,M}p(Y|X,Z,M)\Bigr(\mathbf{1} (M=1)p(Z|X,M = 1)p(M=1) \,+ \\
&\ \mathbf{1} (M=0)p(Z|M = 0)p(M=0)\Bigl) \,,
\end{split} \label{ex_graphs_a_formula}
\end{align}

where the indicator function $\mathbf{1}$ tells in which context $M$, we are calculating the conditional distribution of $Z$ in Equation\eqref{ex_graphs_a_formula}. 

Equation \eqref{ex_graphs_a_formula} can be simplified if we consider one context at a time. In context $M=1$, one can marginalize out $Z$, which serves as a mediator between $X$ and $Y$, and thus should not be conditioned on. In context $M=0$ conditioning on $Z$ is needed to block the backdoor path $X \leftarrow U \rightarrow Z \rightarrow Y$. Thus, an alternative presentation for equation \eqref{ex_graphs_a_formula} can be expressed as

\begin{align}
\begin{split}
P(Y \cond \doo(X)) = & p(Y|X,M=1)p(M=1) \,+ \\
&  \sum_{Z}p(Y|X,Z,M=0)p(Z|M = 0)p(M=0),
\end{split} \label{eq:ex_graphs_alt} 
\end{align}

In the case of LDAG in Figure \ref{fig:ex_graphs}, the result of equation~\eqref{eq:ex_graphs_alt} could be obtained also by applying an identification algorithm separately to contexts $M=0$ and $M=1$ and combining then the identifying functionals. However, in general, the identifiability in all contexts is neither necessary nor sufficient condition for the overall identifiability \citep[Theorem 6]{NEURIPS2019d88518ac}. 

\section{Data and assumptions} \label{data}
\subsection{Survey data} \label{surveydata}

We use open data from Visit Finland's International visitor survey (``Rajahaastattelut'' in Finnish) which consists of interviews of international travellers departing from Finland \citep{avoindata}. Visit Finland, a unit of Business Finland,  is a public organization which builds the brand of Finland as an international destination and helps Finnish companies to develop their travel business internationally \citep{VisitFinlandTehtava}. Statistics Finland, Norstat Finland Oy, and Visitory Oy are responsible for the data collection procedure including survey planning, interviews, data processing, and data publishing \citep{matkailijamittari, OksaNurmi}. The survey including sampling plan, questionnaire, and the data processing is described in \citet{OksaNurmi}. The open data set and detailed documentation of the variables and the survey process are available at \url{https://www.avoindata.fi/data/en_GB/dataset/visit-finland-matkailijamittari} \citep{avoindata}. The data used in our study covers observations from March 2023 to June 2025 and it has been collected on a monthly basis with interviews conducted at various border crossing points in Finland. We next describe the data processing and the post-processed data. The descriptive statistics for the post-processed data sets are provided in Tables \ref{tab:con_vars_M0}, \ref{tab:con_vars_M1}, \ref{tab:cat_vars_M0}, and \ref{tab:cat_vars_M1} in the Appendix \ref{app_data}.

Foreign travellers are defined to be those who are not residing in Finland \citep{matkailijamittari}. In the survey, the interviewer records the traveller's responses using a tablet device \citep{matkailijamittari}. According to \cite{OksaNurmi}, an interviewee can report the expenditure in the chosen currency and for the chosen group of persons. The question was formulated as follows: ``How much money did you spend in Finland? Currency, for the whole party just mentioned. If you do not know, leave the field blank. If there were no expenses, enter 0.'' The reported total expenditure is transformed into euros and then normalized to represent one person's expenditure per one trip \citep{OksaNurmi}.

In addition to expenditure, background variables age, gender, and country of residence have been measured. Due to the low number of observations in the gender variable categories 'Other' and 'Don’t want to say', we merged them with the 'Female' category to create a binary gender variable. The purpose of the trip has seven classes which are divided into two contexts: "Vacation, leisure, or recreation" and "Some other purpose" are categorized as personal trips, and "Studying", "Meeting or work trip in the service of a non-Finnish employer", "Conference or congress or fair", "Work performed in Finland for a Finnish employer", and "Some other work-related reason" are categorized as work-related trips. In the original data study trips were categorized as personal trips but we categorize them as work-related trips category because we assume that the decision-making mechanism for study trips shares the essential features with work-related trips.

The length of stay (nights) is the variable whose causal effect on expenditure will be analysed.
In Table \ref{tab:LOS-table}, the distribution of overnight stays is presented for the both contexts. The majority of trips have the duration of four nights or less but personal trips have a spike also at seven nights.

\begin{table}[ht!!]
\centering\caption{Distribution of the length of stays for personal and work-related trips.}
\begingroup\scriptsize
\begin{tabular}{ccccc}
\\
  \hline
Number of overnight stays & Personal trips (freq.) & \% & Work-related trips (freq.) & \% \\ 
  \hline
1 & 895 & 11.8 & 531 & 15.7 \\ 
  2 & 1225 & 16.2 & 752 & 22.3 \\ 
  3 & 1325 & 17.5 & 557 & 16.5 \\ 
  4 & 944 & 12.4 & 450 & 13.3 \\ 
  5 & 604 & 8.0 & 273 & 8.1 \\ 
  6 & 469 & 6.2 & 168 & 5.0 \\ 
  7 & 1265 & 16.7 & 150 & 4.4 \\ 
  8 & 212 & 2.8 & 85 & 2.5 \\ 
  9 & 176 & 2.3 & 78 & 2.3 \\ 
  10 & 113 & 1.5 & 66 & 2.0 \\ 
  11 & 91 & 1.2 & 70 & 2.1 \\ 
  12 & 70 & 0.9 & 41 & 1.2 \\ 
  13 & 57 & 0.8 & 29 & 0.9 \\ 
  14 & 85 & 1.1 & 85 & 2.5 \\ 
  15 & 54 & 0.7 & 41 & 1.2 \\ 
   \hline
\end{tabular}
\endgroup 
\label{tab:LOS-table}
\end{table}

Other trip-related factors have also been measured, such as the main destinations, and the time of the trip. As the Finland's tourism is significantly concentrated to Helsinki and Lapland, we use the major regions of main destination in Finland indicating the main destination of the tourists. The major regions are Helsinki metropolitan area, Coast and archipelago, Lakeland, and Lapland. In addition, Helsinki is separated from Helsinki metropolitan area as an own region in our dataset, making the final classification of the variable to consider five regions. For the time of the trip, we use the quarter of departure from Finland to represent the traveller's visiting period in Finland.

The travellers interviewed were also asked to report up to three secondary destinations and the length of stay at these. Given the substantial proportion of missing observations in the reported secondary destinations, we adopt the assumption that missing responses indicate the absence of secondary visits. Furthermore, due to low frequency of visits to secondary destinations, this information was summarized as the total sum of overnight stays in the secondary destinations. 

The travellers were asked about the main accommodation and the travel group. Due to relatively low number of observations in some categories of the accommodation variable, we keep the three categories with the highest number of observations and combine the rest of the categories into a single category labelled 'Other'. This procedure was applied separately to the data sets of both contexts. 

In addition, dichotomous questions have been used for different experiences and activities, such as nature or cultural experiences, specifically indicating whether a traveller had participated in such an experience or activity. In these, the travellers interviewed were allowed to report three activities at a maximum. In addition, the first reservation made in months since the trip occurred was used in the analysis. Specific questions were also asked about transportation, considering whether there were trips longer than 50~kilometres within Finland, and the mean of transport when leaving Finland.

As a part of data, Statistics Finland provides sampling weights that are needed, for instance, to estimate the total national income from foreign travellers. Statistics Finland has also imputed some of the variables in the data set. The imputed variables include six sub-expenditures (e.g. total expenditures for accommodation, restaurants, fuels, etc.), that is, the costs that form the total expenditure as well as the length of stay and the purpose of the trip. According to \citet{OksaNurmi}, missing data were imputed using the nearest neighbour method. From the original data dimension ($n$=17\,297) the number of imputed values for the length of stay and the purpose of the trip consider only 52 and 40 observations respectively. However, the share per sub-expenditure variables missingness varies from 12\% to 38\% .

Not all missing values in data are imputed by Statistics Finland. 
In the original data, the explanatory variables containing lots of missing values, such as information about package trips and the number of children participating in the trip have been excluded from the data. Empty values or unknown values reported in the main destination variable are treated as missing data.

Accommodation services constitute a major component of travellers' expenses with a possible exception of travellers who meet their friends and relatives. Thus, we focus on personal and work-related trips excluding the trips made for meeting friends or relatives. We restrict to travellers whose length of stay is between 1 to 15 nights and whose expenditure is greater than zero. Studying the effect of length of stay for travellers with zero expenditure or large number of overnight stays would be atypical behaviour from mainstream and out of our scope. Although daily travellers would be a relevant segment to include in the study, their characteristics in the decision-making process would differ significantly from the overnight stay visitors, e.g. excluding the decision related to accommodation. The processed data sets contain 7585 observations in personal trip context data and 3376 observations in the work-related trip context data. In the final data sets used in fitting the models in Section \ref{estimation}, rows containing missing values are excluded, which reduces the number of observations from 7585 to 6894 in the context of personal trips and from 3376 to 3267 in the context of work-related trips.

As described by \citet{OksaNurmi}, the survey has some limitations related to the data collection process. Road border crossing points were excluded, which reduces the number of Swedish and Norwegian travellers in the data. The costs covered by employer seem to be reported inconsistently and the real expenditures of work related trips are likely to be higher than those in the data.

\subsection{Causal model for traveller's total expenditure}\label{LDAG}

Next, we discuss the causal mechanisms related to the variables in the border survey data and present the assumed causal model as an LDAG. In Table \ref{tab:vars}, we have listed the symbols for the variables used when constructing the causal graph. 

Our context variable $M$ is the purpose of the trip considering 'personal trips' ($M=0$) and 'work-related trips' ($M=1$). These two contexts allow us to expand the description of the data-generating process, leading to distinct independence structures between variables in the causal graph.
In personal trips, a diverse array of motives to travel can be assumed, rendering it ambiguous as to which factors drive the trip planning process. On the other hand, in the case of work-related trips, the work assignments guide the planning.

\begin{table}[ht!!]
    \centering
    \caption{Variables and their definitions.}
    \begin{tabular}{cl}
      \\
  \hline
    Variable  &  Definition \\ \hline
       $D$  &  \textbullet \, Age group, gender, and the country of residence\\ 
       $M$  &  \textbullet \, Purpose of the trip (personal or work-related)\\
       $X$  &  \textbullet \, Length of stay (in number of overnight stays)\\
       $C$  &  \textbullet \, Main destination (Major region of main destination in Finland)\\
       $S$  &  \textbullet \, Quarter of departure from Finland\\ 
 $Z$ & Other decisions assumed to be made before the trip\\
     & \textbullet \, Means of transport for departure from Finland\\
     & \textbullet \, Main type of accommodation\\
     & \textbullet \, How many months prior to the trip was the first reservation made\\
     & \textbullet \, Travel group\\
     & \textbullet \, Overnight stays in other destinations in Finland\\ 
 $W$ & Other decisions assumed to be made before or during the trip\\
     & \textbullet \, (Did or experienced) nature experiences in Finland\\
     & \textbullet \, (Did or experienced) physical or sport activities outdoors in Finland\\
     & \textbullet \, (Did or experienced) wellbeing and relaxation activities in Finland\\
     & \textbullet \, (Did or experienced) cultural experiences in Finland\\
     & \textbullet \, (Did or experienced) a city break in Finland\\
     & \textbullet \, (Did or experienced) participation in a cultural or sport event in Finland\\
     & \textbullet \, (Did or experienced) shopping in Finland\\
     & \textbullet \, (Did or experienced) touring or road trip in Finland\\
     & \textbullet \, Travelled more than 50 kilometres (30 miles) in Finland\\ 
       $Y$  &  \textbullet \, Total expenditure \\ \hline
    \end{tabular}
    \label{tab:vars}
\end{table}

The key variables in the data related to trip planning are denoted as follows: Variable $X$ is the trip length in number of nights. Variables $C$ and $S$ indicate the main destination (major region of main destination) and the season the trip was made (departure time), respectively. $Z$ represents additional characteristics of the trip that are assumed to have been determined before the trip, that is, months the first reservation was made prior to the trip, the mean of transport from Finland, the main type of accommodation, overnight stays in secondary destinations and the travel group. Additional factors that are assumed to have been decided before or during the trip are included in variable $W$, that is, different experiences or activities done and trips over 50 kilometres in Finland. Variables $D$ (age, gender, country of residence), $Z$ and $W$ are clustered variables and are assumed to represent transit clusters \citep{tikka2023clustering}.

In the context of personal trips $M=0$, there exists a broad degree of flexibility in the trip planning, and we assume a common unobserved confounder $U_1$ between all the factors related to the decisions $X$, $C$, $S$, $Z$, and $W$ described above. This $U_1$ could represent, for instance, different types of consumer preferences or lifestyles, advertising, or recommendations from other travellers. Additional unmeasured confounder $U_2$ between key factors $X$, $C$, and $S$ could include other unmeasured confounders, such as constraints related to timing and scheduling.

In the context of work-related trips $M=1$, we assume that the requirements of work or studies are essentially guiding the choose between primary factors, that is, the destination and timing (season and length of stay). This means that the unobserved $U_1$ (e.g. consumer preferences) is no longer affecting on the choice of $X$, $C$, $S$. However, the effect of $U_1$ is present between $Z$ and $W$, reflecting the influence of $U_1$ on the consumption of services and activities. Besides the aforementioned traveller's constraints in personal trip contexts, the unmeasured confounder $U_2$ between key factors $X$, $C$, and $S$ could here include e.g. additional information about the requirements and features of the work or studies. The key factors $X$, $C$, and $S$ together further affect the rest of the decisions related to the trip, that is, $Z$ and $W$.  

\begin{figure}[ht!!]
\centering
\scriptsize
\begin{tikzpicture}[>=Latex, fill=white, scale=0.75]

\node[draw, fill=white, mycircle] at (-5,4) (D) {$D$};
\node[draw, fill=white, myrectangle] at (5,4) (U1) {$U_1$};
\node[draw, fill=white, mycircle] at (0,4) (M) {$M$};
  
\node[draw, fill=white, mycircle] at (0,-7) (Y) {$Y$};

\node[draw, fill=white, mycircle] at (-5,0) (X) {$X$};
\node[draw, fill=white, mycircle] at (0,0) (C) {$C$};
\node[draw, fill=white, mycircle] at (5,0) (S) {$S$};
\node[draw, fill=white, myrectangle] at (-7.5,4) (U2) {$U_2$};
\node[draw, fill=white, mycircle] at (2,-3) (Z) {$Z$};
\node[draw, fill=white, mycircle] at (-2,-5) (W) {$W$};

\draw[->] (D) -- (X);
\draw[->] (D) -- (C);
\draw[->] (D) -- (S);
\draw[->] (D) -- (Z);
\draw[->] (D) -- (W);
\draw[->] (D) -- (M);
\draw[->] (D) -- (Y);

\draw[->] (M) -- (X);
\draw[->] (M) -- (C);
\draw[->] (M) -- (S);
\draw[->] (M) -- (Z);
\draw[->] (M) -- (W);
\draw[->] (M) to [bend left=60] (Y);

\draw[->] (Z) -- (Y);
\draw[->] (Z) -- (W);
\draw[->] (W) -- (Y);

\draw[->] (U1) -- (Z);
\draw[->] (U1) -- (W);

\draw[->] (S) -- (Y);

\draw[->] (X) to [bend right=40] (Y);

\draw[->] (C) -- (Y);

\draw[dashed, ->, gray!250] (X) -- node[midway, above, text opacity=1, fill=white, font=\scriptsize] {$M=0$} (Z);
\draw[dashed, ->, gray!250] (X) -- node[midway, above, text opacity=1, fill=white, font=\scriptsize] {$M=0$} (W);

\draw[dashed, ->, gray!250] (C) -- node[midway, above, text opacity=1, fill=white, font=\scriptsize] {$M=0$} (Z);
\draw[dashed, ->, gray!250] (C) -- node[midway, above, text opacity=1, fill=white, font=\scriptsize] {$M=0$} (W);

\draw[dashed, ->, gray!250] (S) -- node[midway, above, text opacity=1, fill=white, font=\scriptsize] {$M=0$} (Z);
\draw[dashed, ->, gray!250] (S) to [bend left=60] node[midway, above, text opacity=1, fill=white, font=\scriptsize] {$M=0$} (W);

\draw[->] (U2) -- (X);
\draw[->] (U2) to [bend left=60] (S);
\draw[->] (U2) -- (C);

\draw[dashed, ->, gray!250] (U1) -- node[near start, above, text opacity=1, fill=white, font=\scriptsize] {$M=1$} (X);
\draw[dashed, ->, gray!250] (U1) -- node[near start, above, text opacity=1, fill=white, font=\scriptsize] {$M=1$} (C);
\draw[dashed, ->,  gray!250] (U1) -- node[near start, above, text opacity=1, fill=white, font=\scriptsize] {$M=1$} (S);

\end{tikzpicture}
\caption{\label{fig:LDAG}LDAG representing traveller's total expenditure process: The circle nodes represent the observed variables in the data. The labels in the arrows, describe the causal paths that vanish when context is either personal trip ($M=0$), or work-related trip ($M=1$). The rectangle nodes $U_1$ and $U_2$ indicate unobserved confounders.}
\end{figure}
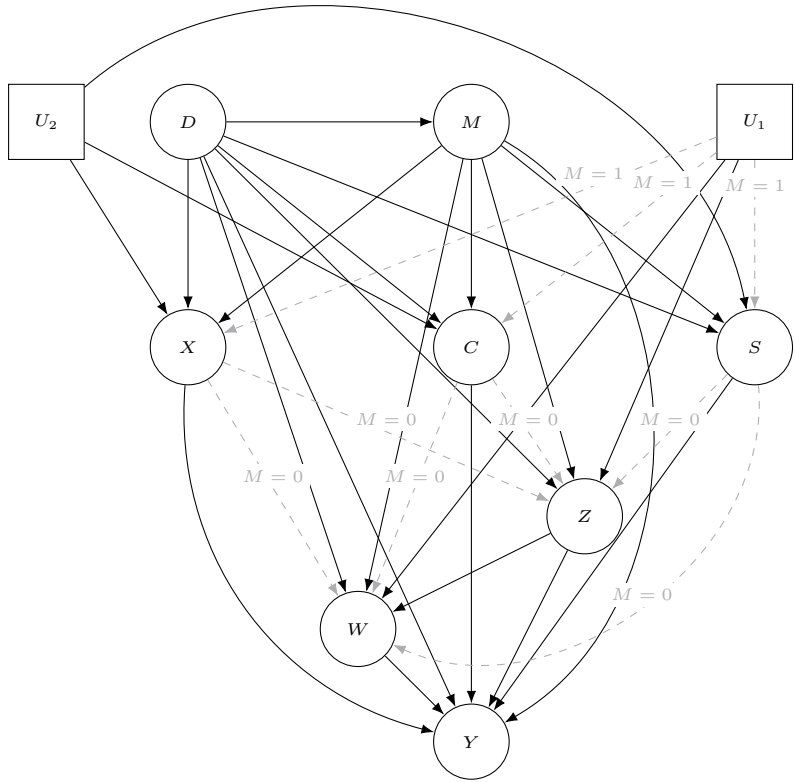

To combine the information from both of the contexts together into LDAG presented in Figure~\ref{fig:LDAG}, we first describe the rest of the assumed causal relations between data variables. In the LDAG, $D$ is assumed to affect directly the purpose of the trip $M$, characteristics $X$, $C$, $S$, $Z$, and $W$, and the total expenditure $Y$. The purpose of the trip $M$ (the context variable) is a confounder between total expenditure $Y$ and all the factors concerning the decisions. In addition, all of these variables directly affect to the total expenditure $Y$. By using LDAG, the differences in the causal mechanisms due the contexts are easy to present and thus the whole data generating process is described in one graph instead of multiple DAGs.

\begin{figure}[ht!!]
\centering
\scriptsize
\begin{tikzpicture}[>=Latex, fill=white, scale=0.75]

\node[draw, fill=white, mycircle] at (-5,4) (D) {$D$};
\node[draw, fill=white, myrectangle] at (5,4) (U1) {$U_1$};
\node[draw, fill=white, mycircle] at (0,4) (M) {$M$};
  
\node[draw, fill=white, mycircle] at (0,-7) (Y) {$Y$};

\node[draw, fill=white, mycircle] at (-5,0) (X) {$X$};
\node[draw, fill=white, mycircle] at (0,0) (C) {$C$};
\node[draw, fill=white, mycircle] at (5,0) (S) {$S$};
\node[draw, fill=white, myrectangle] at (-7.5,4) (U2) {$U_2$};
\node[draw, fill=white, mycircle] at (2,-3) (Z) {$Z$};
\node[draw, fill=white, mycircle] at (-2,-5) (W) {$W$};

\draw[->, lightgray] (D) -- (X);
\draw[->, lightgray] (D) -- (C);
\draw[->, lightgray] (D) -- (S);
\draw[->, lightgray] (D) -- (Z);
\draw[->, lightgray] (D) -- (W);
\draw[->, lightgray] (D) -- (M);
\draw[->, lightgray] (D) -- (Y);

\draw[->, lightgray] (M) -- (X);
\draw[->, lightgray] (M) -- (C);
\draw[->, lightgray] (M) -- (S);
\draw[->, lightgray] (M) -- (Z);
\draw[->, lightgray] (M) -- (W);
\draw[->, lightgray] (M) to [bend left=60] (Y);

\draw[->, lightgray] (Z) -- (Y);
\draw[->, lightgray] (Z) -- (W);
\draw[->, lightgray] (W) -- (Y);

\draw[->, lightgray] (U1) -- (Z);
\draw[->, lightgray] (U1) -- (W);

\draw[->, lightgray] (S) -- (Y);

\draw[->, lightgray] (X) to [bend right=40] (Y);

\draw[->, lightgray] (C) -- (Y);

\draw[dashed, ->, lightgray] (X) -- node[midway, above, text opacity=1, fill=white, font=\scriptsize] {$M=0$} (Z);
\draw[dashed, ->, lightgray] (X) -- node[midway, above, text opacity=1, fill=white, font=\scriptsize] {$M=0$} (W);

\draw[dashed, ->, lightgray] (C) -- node[midway, above, text opacity=1, fill=white, font=\scriptsize] {$M=0$} (Z);
\draw[dashed, ->, lightgray] (C) -- node[midway, above, text opacity=1, fill=white, font=\scriptsize] {$M=0$} (W);

\draw[dashed, ->, lightgray] (S) -- node[midway, above, text opacity=1, fill=white, font=\scriptsize] {$M=0$} (Z);
\draw[dashed, ->, lightgray] (S) to [bend left=60] node[midway, above, text opacity=1, fill=white, font=\scriptsize] {$M=0$} (W);

\draw[->, lightgray] (U2) -- (X);
\draw[->, lightgray] (U2) to [bend left=60] (S);
\draw[->, lightgray] (U2) -- (C);

\draw[dashed, ->, lightgray] (U1) -- node[near start, above, text opacity=1, fill=white, font=\scriptsize] {$M=1$} (X);
\draw[dashed, ->, lightgray] (U1) -- node[near start, above, text opacity=1, fill=white, font=\scriptsize] {$M=1$} (C);
\draw[dashed, ->, lightgray] (U1) -- node[near start, above, text opacity=1, fill=white, font=\scriptsize] {$M=1$} (S);


\node[draw, fill=white, myrectangle] at (-8.5,-5) (I) {$I$};

\draw[->] (D) -- (I);
\draw[->] (I) -- (X);
\draw[->] (I) -- (C);
\draw[->] (I) -- (S);
\draw[->] (I) -- (Z);
\draw[->] (I) -- (W);
\draw[->] (I) -- (M);
\draw[->] (I) -- (Y);

\end{tikzpicture}
\caption{\label{fig:LDAG_augmented}An augmented LDAG, where traveller's income $I$ introduces new causal links including a confounding pathway between the length of stay $X$ and traveller's total expenditure $Y$.}
\end{figure}
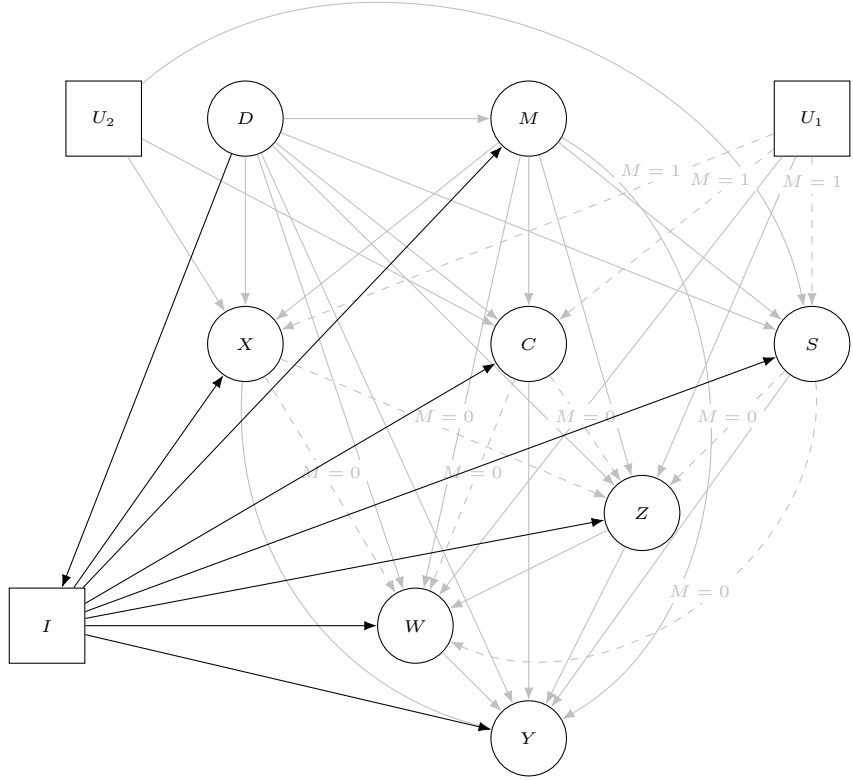

The data lacks some key background variables of travellers considering traveller's financial resources, such as income. In Figure \ref{fig:LDAG_augmented}, we present an augmented LDAG incorporating traveller's income in the process. It is plausible to assume, that traveller's income influences both the length of stay and total expenditure, thus preventing the identification and introducing bias in the estimates if not accounted for. To address this issue, we implement a sensitivity analysis in Section \ref{ovb}. 
 
\newpage
\section{Identification of causal effects and counterfactuals}
\label{sec_identif}

In this section, we first consider the identification of causal effects in the LDAGs presented in Figures~\ref{fig:LDAG} and \ref{fig:LDAG_augmented}. Then we study the counterfactual situation where travellers would have stayed one night longer than they in reality did.

We are interested in the post-interventional distribution $P(Y \cond \doo(X))$ that describes the causal effect of length of stay ($X$) on the total expenditure ($Y$). 
Without CSI-relations the graph of Figure \ref{fig:LDAG}  reduces  to a regular DAG, where the context-specific edges are always present. In this case, the query $P(Y \cond \doo(X))$ is not identifiable. With CSI-relations, $P(Y \cond \doo(X))$ is identifiable from the observational distribution $P(Y,X,C,S,Z,W,D,M)$ and the Do-search algorithm \citep{NEURIPS2019d88518ac,dosearchtikka2021} returns an identifying functional 

\begin{align}
\footnotesize
\label{y_dox}
\begin{split}
P(Y \cond \doo(X)) &= \sum_{M,\mathcal{S}_0}\Bigr(\mathbf{1}(M=0)p(M)p(\mathcal{S}_0|M)p(Y|X,M,\mathcal{S}_0)\Bigl) + \\
&\Bigr(\mathbf{1}(M=1)p(Y|X,M,\mathcal{S}_0)p(W|X,M,\mathcal{S}_2)p(Z|X,M,\mathcal{S}_1)p(M,\mathcal{S}_1)\Bigl),
\end{split}
\end{align}
where $\mathcal{S}_0=\{C,S,Z,W,D\}$ and $\mathcal{S}_1=\{C,S,D\}$, and $\mathcal{S}_2=\{C,S,Z,D\}$.
The causal effects conditioned on the context  $P(Y \cond \doo(X),M)$ are identified as follows

 \begin{equation}
 \footnotesize
\begin{split}
 \label{eq:y_dox_m_M0}
 P(Y \cond \doo(X),M=0) = 
 \frac{p(M=0)\sum\limits_{\mathcal{S}_0}p(\mathcal{S}_0|M=0)p(Y|X,M=0,\mathcal{S}_0)}{\sum\limits_{Y}\Bigr[p(M=0)\sum\limits_{\mathcal{S}_0}p(\mathcal{S}_0|M=0)p(Y|X,M=0,\mathcal{S}_0)\Bigl]}   
 \end{split}
 \end{equation}

 \begin{equation}
 \footnotesize
\begin{split}
 \label{eq:y_dox_m_M1}
 P(Y \cond \doo(X),M=1) = 
 \frac{\sum\limits_{\mathcal{S}_1}p(Y|X,M=1,\mathcal{S}_1)p(M=1,\mathcal{S}_1)}{\sum\limits_{Y}\Bigr[\sum\limits_{\mathcal{S}_1}p(Y|X,M=1,\mathcal{S}_1)p(M=1,\mathcal{S}_1)\Bigl]}\,. 
\end{split}
 \end{equation}

In the above, the identifiability of $P(Y \cond \doo(X),M)$ could also be reached by presenting the CSI-relations separately with DAGs for both contexts. From a conceptual standpoint, $P(Y \cond \doo(X),M)$ means intervening directly the traveller's length of stay, i.e. setting the overnight stay as $X=x$. 

The main question of interest is finding  the traveller's total expenditure if he/she had stayed one night longer than originally planned. The estimand of interest is then a counterfactual difference which can be defined separately for both contexts as

\begin{align}
\label{cf_eq}
\Delta_{x \rightarrow x+1}(x,m) &:= \E(Y_{x+1}|X=x,M=m)-\E(Y_x|X=x,M=m),
\end{align}

where $\E(Y_x|X=x,M=m)$ is the expected total expenditure for travellers who stayed $x$ nights and $\E(Y_{x+1}|X=x,M=m)$ is the counterfactual total expenditure for the same travellers if they had stayed one night longer. The identifiability of counterfactuals can evaluated by applying the algorithm by \citet{Shpitser2007counterfactuals} implemented in the R package \texttt{cfid} \citep{tikka2023cfid}. The following formulas can be derived from the identifying functionals:

\begin{subequations}\label{cf_expected_values}
\begin{align}
   & \E(Y_{x}|X=x,M=0) = \frac{P(\mathcal{S}_0 \cond X=x, M=0)\E(Y \cond X=x,\mathcal{S}_0,M=0)}{P(X=x \cond M=0)}, \label{eq:cf_x_M0}\\
   & \E(Y_{x}|X=x,M=1) = \frac{P(\mathcal{S}_1 \cond X=x, M=1)\E(Y \cond X=x,\mathcal{S}_1,M=1)}{P(X=x \cond M=1)}, \label{eq:cf_x_M1}\\
   & \E(Y_{x+1}|X=x,M=0) = \frac{P(\mathcal{S}_0 \cond X=x, M=0)\E(Y \cond X=x+1,\mathcal{S}_0,M=0)}{P(X=x \cond M=0)}, \label{eq:cf_x1_M0}\\
   & \E(Y_{x+1}|X=x,M=1) = \frac{P(\mathcal{S}_1 \cond X=x, M=1)\E(Y \cond X=x+1,\mathcal{S}_1,M=1)}{P(X=x \cond M=1)},\label{eq:cf_x1_M1}
\end{align}
\end{subequations}
where again $\mathcal{S}_0=\{C,S,Z,W,D\}$ and $\mathcal{S}_1=\{C,S,D\}$. Note that in equations~\eqref{eq:cf_x1_M0} and \eqref{eq:cf_x1_M1}, the distributions of $\mathcal{S}_0$ and $\mathcal{S}_1$ are conditional on $X=x$ but the distribution of $Y$ is conditional on $X=x+1$.

The counterfactual
\[
\E(Y_{x+1}|Y=y,X=x,M=m),
\]
where the condition for $Y_{x+1}$ contains also the actual expenses $y$ would be interesting as well. However, this counterfactual is not identifiable.

\section{Estimating the effect of extending the length of stay}
\label{estimation}

We estimate the effect of extending the length of stay on total expenditure, i.e., the counterfactual difference $\Delta_{x \rightarrow x+1}(x,m)$ in Equation \eqref{cf_eq}, in both contexts $m=\{0,1\}$. To model both expected total expenditure and counterfactual total expenditure in Equation \eqref{cf_eq}, we need a model for the expected value $\E(Y \cond X,M)$ for both contexts. 

The equations in \eqref{eq:cf_x_M0}-\eqref{eq:cf_x1_M1} tell that we specifically need to model the conditional expectations $\E(Y \cond X,\mathcal{S}_0,M=0)$ and $\E(Y \cond X,\mathcal{S}_1,M=1)$. As we have restricted the total expenditure to consider only positive values, we fit gamma distribution models. We apply logarithmic link functions and allow the intercepts to vary by incorporating random effects for the country of residence ($L$) of a traveller. We assume the effect of increasing the length of stay on the expenditure to be non-decreasing, and therefore monotonic effects for the length of stay \citep{burknermonotonic} are estimated.

Let the superscripts define the context related to the variables and parameters, that is, zero for context $M=0$ and one for context $M=1$. The models for the the total expenditure $Y$ of a traveller $i$ with country of residence $L[i]$ can be expressed as

\begin{align}
\begin{split} \label{model_M0}
& \log(\E(Y_i \cond X_i=x,\mathcal{S}_0,M_i=0)) = \\
& \alpha^{(0)} + \bm{\beta}^{(0)}_{M^{(0)}[i]} +
\bm{\gamma}^{(0)}\mathcal{S}^{*}_{0i} + u^{(0)}_{L[i]} + a^{(0)}\moo(x,\bm{\zeta}^{(0)}),
\end{split}
\end{align}

\begin{align}
\begin{split} \label{model_M1}
& \log(\E(Y_i \cond X_i=x,\mathcal{S}_0,M_i=1)) = \\ 
& \alpha^{(1)} + \bm{\beta}^{(1)}_{M^{(1)}[i]} +
\bm{\gamma}^{(1)}\mathcal{S}^{*}_{1i} + u^{(1)}_{L[i]} + a^{(1)}\moo(x,\bm{\zeta}^{(1)}),
\end{split}
\end{align}

where $\alpha$ is a constant, $\bm{\beta}$ is the parameters regarding the different purposes of travel in each context, and $\bm{\gamma}$ refers to parameters related to other covariates in the specific context. Parameter $a$ defines the direction and the magnitude of the monotonic effect and $\moo$ function is the monotonic transform with simplex parameters $\bm{\zeta}$ \citep{burknermonotonic}.

Random intercept $u_{L[i]}$ model the effect of the $i$th traveller's country of residence $L$. In the formulas, variables $M^{(0)}$ and $M^{(1)}$ now include the sub-categories of the purpose of the travel in the corresponding contexts, presented in Section \ref{data}. Note also, that sets $\mathcal{S}^{*}_{0i}$ and $\mathcal{S}^{*}_{1i}$ include the variables from sets $\mathcal{S}_0$ and $\mathcal{S}_1$ correspondingly, excluding the country of residence already presented in random effect $u_{L[i]}j$. In sets $\mathcal{S}^{*}_{0i}$ and $\mathcal{S}^{*}_{1i}$, categorical variables are presented by binary indicators for each class.

The model is implemented in the R environment \citep{r} by using the \texttt{brms} package \citep{brms1} for Bayesian regression modelling. Weakly informative priors are used in the modelling. We choose $N(0,0.5^2)$ priors for regression coefficients, and $Dirichlet(2)$ priors for the concentration parameters (simplex). For intercept term of the mean parameter, $N(0,2^2)$ prior is used, and $t(3,0,1)$ prior for the standard deviation, with 3 degrees of freedom, 0 location and 1 scale parameter. Otherwise, the default priors from the \texttt{brms} package are used. We use 2000 iterations with 1000 warm-up period, and run 4 chains.

The expected values of the posterior predictive distribution were scaled with the sampling weights provided by Statistics Finland. For estimating the effect of extending the length of stay, we calculate the counterfactual difference in Equation  \eqref{cf_eq} for all posterior samples $b=1,...,B$, where $\boldsymbol{\theta}_b$ stands for all model parameters for sample $b$. The expected counterfactual difference of increasing the length of stay on the total expenditure by one night is estimated by the mean

\begin{align}
\label{cf_eff}
\widehat{\Delta}_{x \rightarrow x+1}(x,m) &= \frac{1}{B}\sum_{b=1}^{B}\Delta_{x \rightarrow x+1}(x,m;\boldsymbol{\theta}_b),
\end{align}

where $b$ represents the posterior sample and $B$ the total number of posterior samples. 95\% credible interval (CI) is obtained from the 2.5th and 97.5th quantiles of $\Delta_{x \rightarrow x+1}(x,m;\boldsymbol{\theta}_b)$. 

\begin{figure}[htbp]
  \centering
  \includegraphics[width=0.8\textwidth]{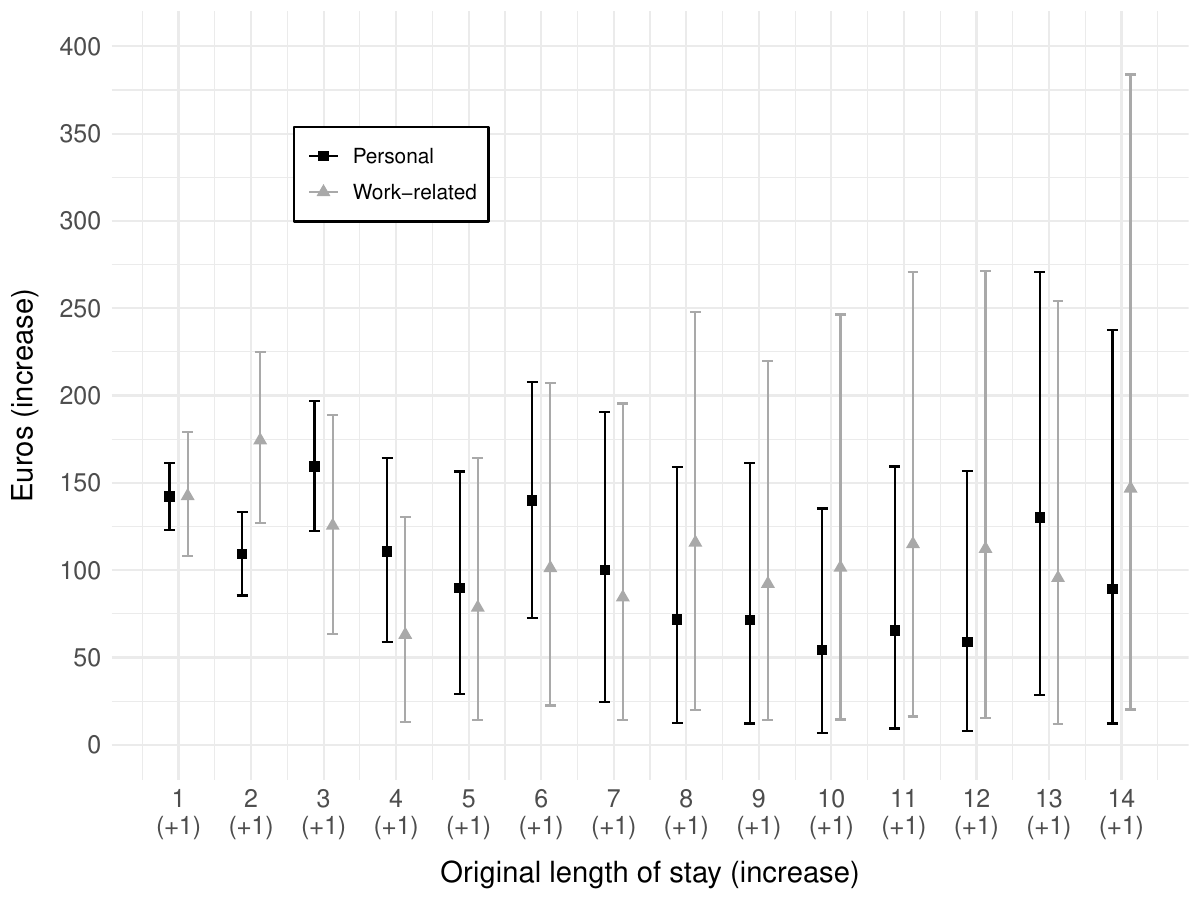}
  \caption{Counterfactual difference of increasing the length of stay by one night for personal trips and work-related trips. Labels on the x-axis indicate the original length of stay and the increment applied (in parenthesis).}
  \label{fig:ace_plot}
\end{figure}

Figure \ref{fig:ace_plot} presents the expected counterfactual difference resulting from increasing the length of stay on total expenditure by one night. Corresponding numerical estimates and their 95 \% credible intervals are presented in Table \ref{tab:ACE} in the Appendix \ref{app_est}.

For personal trips, the increase in total expenditure within one to seven overnight stays varies approximately between 90 to 160 euros. For overnight stays longer than seven nights, the increase in expenditure decreases near to 50 euros, with spike at thirteen nights. However, the credible intervals for the longer length of stays are wide. For work-related trips the highest increase in expenditure is observed at two overnigth stays (approximately 174 euros). After short length of stays, the inrease varies from 60 to 150 euros albeit with substantial uncertainty.

\section{Sensitivity analysis for omitted variable bias}\label{ovb}

Next, we consider the confounding effect of traveller's income which was not asked in the survey. We assume that demographics $D$  affect income $I$, which in turn affects travel-related characteristics $X$, $C$, $S$, $Z$, $W$, and also the purpose of the trip $M$ and the total expenditure $Y$. As it was mentioned in Section~\ref{sec_identif}, adding the income in the graph in Figure\ref{fig:LDAG_augmented} makes the causal effect of interest unidentifiable due to the introduction of additional confounding pathways between $X$ and $Y$. 

To implement a sensitivity analysis of omitted variable bias concerning income, we adapt concepts presented by \citet{cinelli2020making} for Bayesian analysis. We assume a linear relationship between income ($I$) and total expenditure ($Y$), as well as between income ($I$) and the length of stay ($X$). Based on this assumption, we specify prior distributions for the correlations between $I$ and $Y$, and between $I$ and $X$.

We assume that in the context of personal trips, $I$ and $X$ have a weak positive correlation with moderate uncertainty. Although it is plausible that travellers with higher incomes can afford longer trips, it is unclear whether longer stays would be preferred over short visits. For $I$ and $Y$, a moderate correlation is assumed, because travellers typically finance their trip expenses by themselves. 

Also in the context of work-related trips,  $I$ and $X$ are assumed to have a weak positive correlation with moderate uncertainty. This is due to the constraints imposed by work-related obligations which restrict the length of stay even if the traveller could afford a longer stay. The correlation between $I$ and $Y$ is moderate but weaker than in the context of personal trips. Employers subsidize travel costs but an important position at work may imply both high income and high travel costs. Even if employers covered the costs, it is unclear whether travellers differentiate costs unambiguously when reporting them in the survey.

The correlations are simulated from their priors for each posterior sample of the fitted model by using an appropriate normal distributions in combination with the Inverse Fisher $z$-transformation. For personal trips, this leads to prior distribution $N(0.1, 0.15^2)$ for the correlation between $I$ and $X$, and $N(0.45, 0.1^2)$ for the correlation between $I$ and $Y$. For work-related trips, we obtain prior distribution $N(0.05, 0.15^2)$ for the correlation between $I$ and $X$, and $N(0.2, 0.1^2)$ for the correlation between $I$ and $Y$. 

We calculate the bias of the causal effect by applying the following formula \citep{cinelli2020making}:

\begin{align}
\label{bias_formula}
\begin{split}
\widehat{\text{bias}}
=\text{sign}(\text{corr}(Y^{\bot \mathcal{S},X},I^{\bot \mathcal{S},X})\,
\text{corr}(X^{\bot \mathcal{S}}, I^{\bot \mathcal{S}}))
\sqrt{\frac{R^2_{Y \sim I|X,\mathcal{S}}R^2_{X \sim I|\mathcal{S}}}
{1-R^2_{X \sim I|\mathcal{S}}}\frac{\text{sd}(Y^{\bot \mathcal{S},X})}{\text{sd}(X^{\bot \mathcal{S}})}
}\, .
\end{split}
\end{align}
In Equation \eqref{bias_formula} the partial coefficients of determinations $R^2_{Y \sim I|X,\mathcal{S}}$ and $R^2_{X \sim I|\mathcal{S}}$ can be computed from the correlations $\text{corr}(Y^{\bot \mathcal{S},X},I^{\bot \mathcal{S},X})$ and $\text{corr}(X^{\bot \mathcal{S}}, I^{\bot \mathbf{S}})$ generated above, which also defines the sign of the bias. The standard deviation $\text{sd}(Y^{\bot \mathcal{S},X})$
can be calculated from the residuals of the models \eqref{model_M0} and \eqref{model_M1}. To estimate the standard deviations $\text{sd}(X^{\bot \mathcal{S}})$, we
fit  gamma models with logarithmic link functions for the length of stay in both contexts using the same covariate sets in than in the total expenditure models, that is, $\mathcal{S}_0$ for the personal trip context model, and $\mathcal{S}_1$ for the work-related trip context model (excluding the length of stay $X$ which is now the response variable). The models are fitted with prior distributions $N(0,2^2)$ for the regression coefficients, otherwise the default priors are used.

\begin{figure}[htbp!!!]
  \centering
  \includegraphics[width=0.75\textwidth]{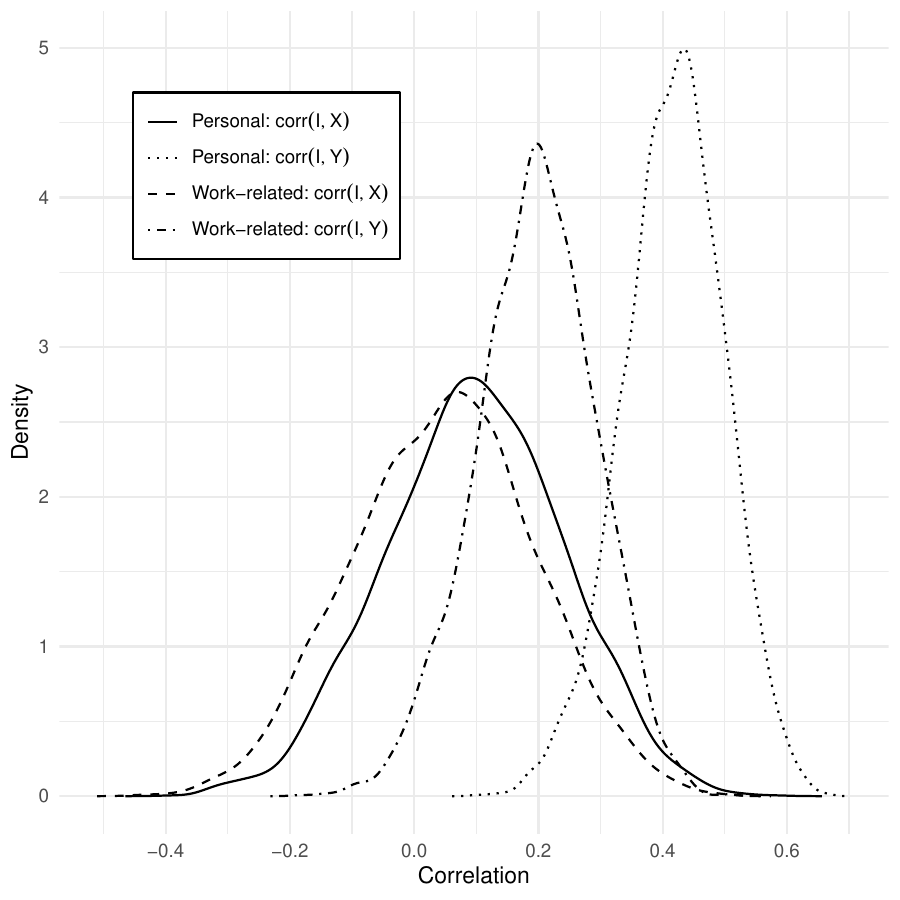}
  \caption{Prior distributions for correlations between income (I) and length of stay (X), and between income and total expenditure (Y) in both personal trip and work-related trip context.}
  \label{fig:corr_prior}
\end{figure}

\begin{figure}[htbp!!!]
  \centering
  \includegraphics[width=0.75\textwidth]{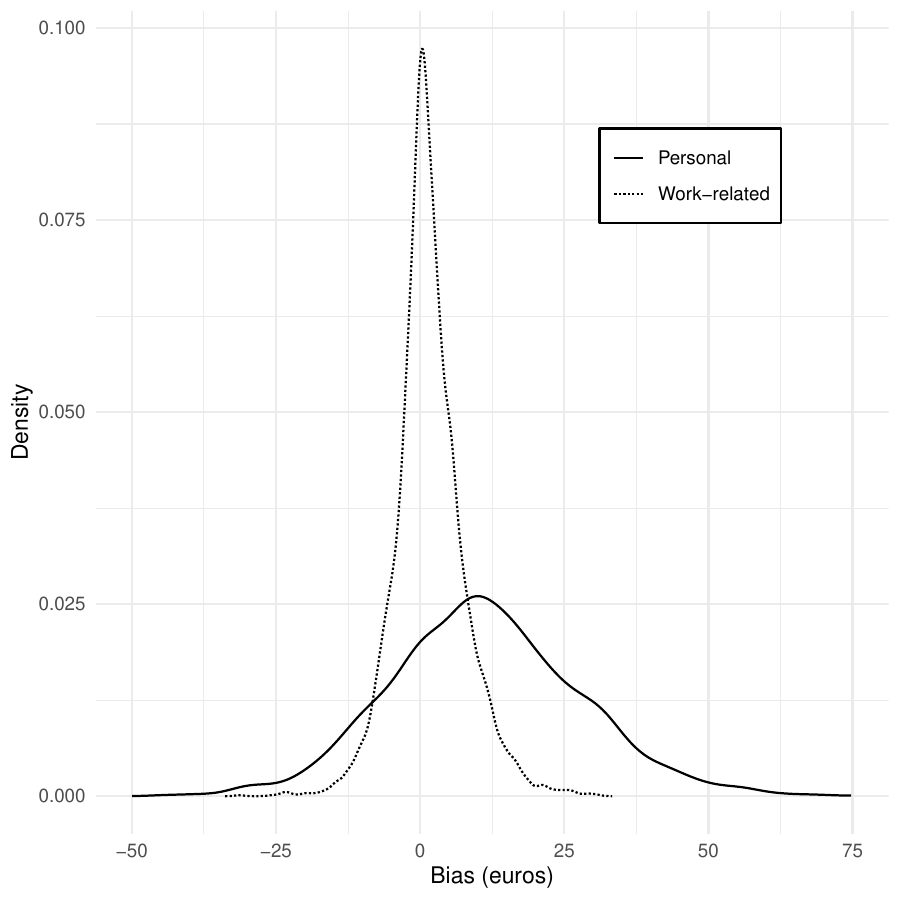}
  \caption{Posterior distributions of omitted variable biases related to missing income variable for personal trips (left) and work-related trips (right).}
  \label{fig:biases}
\end{figure}

\begin{figure}[htbp!!!!]
  \centering
  \includegraphics[width=0.8\textwidth]{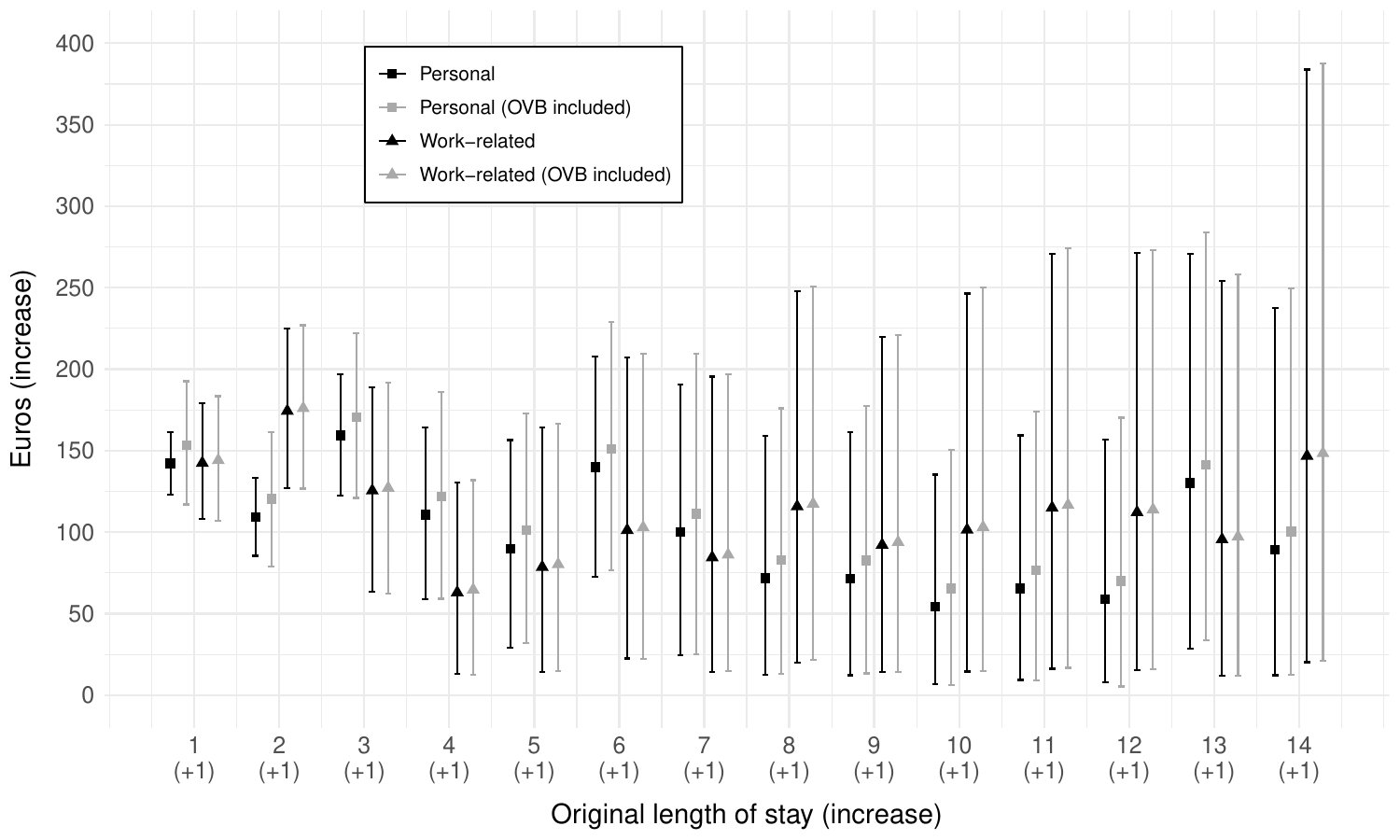}
  \caption{Counterfactual difference of increasing the length of stay by one night on the total expenditure. Models for personal trips and work-related trips excluding and including omitted variable bias related to income are presented. Labels in the x-axis indicates the original length of stay and the increment (in parenthesis).}
  \label{fig:ace_plot_ovb}
\end{figure}

Figure \ref{fig:biases} shows the  bias of the counterfactual difference $\Delta_{x \rightarrow x+1}$ across all posterior samples $\boldsymbol{\theta}_l$ ($l=1,2,...,4\,000$) in both contexts. The estimated average bias of the total expenditure is approximately 11 euros with 95\% quantile interval [-20.00, 45.94] for the posterior distribution for personal trips. For work-related trips, the estimated average bias is approximately 1.9 euros with 95\% quantile interval [-9.91, 15.03] for the posterior distribution. Given the prior belief, omitting income tends to underestimate the bias related to the estimated effect of length of stay in personal trips, although this effect is uncertain with substantial probability mass below zero. For work-related trips, the bias is near zero with high uncertainty. The effect of extending the length of stay on total expenditure in personal trips has tendency to be underestimated if omitting income variable, although the uncertainty related to this is quite large as the probability mass has also lots of mass under zero. The effect of bias on the estimates are presented in Figure \ref{fig:ace_plot_ovb}. The numerical estimates in the omitted variable case are collected in table \ref{tab:ACE_ovb} in the Appendix \ref{app_est}.

\section{Discussion} \label{discussion}

We implemented a context-specific causal model for estimating the counterfactual difference of extending the length of stay on the traveler's total expenditure. To the best of our knowledge, the study presented is the first one to apply context-specific causal inference approach with real data.

LDAGs with CSI-relations enable handling the whole modeling scheme coherently in one causal graph instead of multiple graphs separated by contexts. The approach is theoretically  appealing because identifiability in all context-specific DAGs is neither necessary nor sufficient condition for the identification in LDAG \citep{NEURIPS2019d88518ac}.

The limitations of our analysis are related to data collection, handling of missing data and validity of causal assumptions. The survey design does not include all border crossing points. Recall bias, misunderstood questions due to a lack of a common language, and variability in the reporting of the costs paid by the employer are also potential sources of bias in the data collection. Our analysis focused on short-term personal and work related travel excluding trips where the main purpose was meeting friends and relatives. Modeling of the excluded groups as an additional context would be possible but the main financial interest lies in travelers who use accommodation services.

The missing expenditures have been imputed by Statistics Finland using single imputation. This means that the reported posterior intervals are too narrow because they do not reflect the uncertainty due to missing data. The missing data mechanism could be modeled as a part of the LDAG and included in the identification. Multiple imputation or Bayesian modeling of missing data would be theoretically more justifiable approaches than single imputation, but on the other hand, the imputed data provided by Statistics Finland is used to produce national statistics on international travelers.

Unobserved confounding is a main concern in causal inference with observational data. We addressed this concern through sensitivity analysis examining the impact of missing income data on our results. We devised a Bayesian approach to estimate posterior distributions of the counterfactual expenditure when the correlations between income and the length stay and income and total expenditure followed informative prior distributions. While there remain opportunities to further refine the sensitivity analysis with additional covariates, our approach gives the basic understanding on the extend of omitted variable bias. 

The obtained results pinpoint the overnight stays that generate the highest additional expenditure from an extended visit. This knowledge could be used by destination management organizations to design marketing efforts, such as special offers for extended stay.

\section*{Conflicts of interest}
The authors declare that they have no competing interests.

\section*{Funding}
L.V. was supported by the Foundation for Economic Education. This work was supported by the Research Council of Finland under grant number 368935 and "Enhancing data oriented management of Hämeenlinna city tourism and travelling" project co-funded by the European Union and City of Hämeenlinna.

\section*{Data availability}
The data is available at \url{https://avoindata.suomi.fi/data/en_GB/dataset/visit-finland-matkailijamittari} and the codes are available at \url{https://github.com/lpkvalkonen/cscm}

\section*{Author contributions statement}
The study was designed by both authors. Data analysis was performed by L.V., and the manuscript was written by L.V., and J.K.. Both authors reviewed, read and approved the final manuscript.

\section*{Acknowledgments}
The authors thank Ella Oksa for the aid in data description and Olli Koskela for the comments on the manuscript.


\bibliographystyle{oup-abbrvnat}
\bibliography{bibliography}


\begin{appendices}

\section{Data tables}\label{app_data}

\begin{table}[ht!!]
\centering\caption{Personal trips data ($n$=7585): Summary statistics of continuous variables.} 
\begingroup\tiny
\begin{tabular}{ccccccc}
\\
  \hline
Variable & Mean & Median & SD & Min & Max & Missing Observations \\ 
  \hline
Expenditure (Euros) & 879.04 & 680.00 & 784.68 & 0.24 & 17613.16 & 0 \\ 
  Length of stay (overnight stays) & 4.58 & 4.00 & 2.95 & 1.00 & 15.00 & 0 \\ 
  First reservation (months before the trip) & 3.73 & 2.00 & 4.10 & 0.00 & 40.00 & 393 \\ 
  Overnight stays in the secondary destinations & 0.52 & 0.00 & 1.27 & 0.00 & 12.00 & 152 \\ 
   \hline
\end{tabular}
\endgroup
\label{tab:con_vars_M0}
\end{table}

\begin{table}[ht!!]
\centering\caption{Work-related trips data ($n$=3376): Summary statistics of continuous variables.} 
\begingroup\tiny
\begin{tabular}{ccccccc}
\\
  \hline
Variable & Mean & Median & SD & Min & Max & Missing Observations \\ 
  \hline
Expenditure (Euros) & 652.07 & 487.07 & 645.56 & 0.58 & 10798.64 & 0 \\ 
  Length of stay (overnight stays) & 4.27 & 3.00 & 3.35 & 1.00 & 15.00 & 0 \\ 
  First reservation (months before the trip) & 1.61 & 1.00 & 3.68 & 0.00 & 40.00 & 266 \\ 
  Overnight stays in the secondary destinations & 0.28 & 0.00 & 0.92 & 0.00 & 14.00 & 44 \\ 
   \hline
\end{tabular}
\endgroup
\label{tab:con_vars_M1}
\end{table}

\begin{table}[ht!!]
\centering\caption{Personal trips data ($n$=7585): Frequencies and proportions of categorical variables.} 
\begingroup\tiny
\begin{tabular}{ccccc}
\\
  \hline
Variable & Categories & Frequencies & Proportions (\%) & Missing observations \\ 
  \hline
Main destination & Helsinki & 3294 & 44 & 111 \\ 
   & Helsinki metropolitan area & 117 & 2 &  \\ 
   & Coast and archipelago & 644 & 9 &  \\ 
   & Lakeland & 520 & 7 &  \\ 
   & Lapland & 2899 & 39 &  \\ 
  Quarter & 1 & 2208 & 29 & 0 \\ 
   & 2 & 1806 & 24 &  \\ 
   & 3 & 1753 & 23 &  \\ 
   & 4 & 1818 & 24 &  \\ 
  Gender & Male & 3725 & 50 & 141 \\ 
   & Other than male & 3719 & 50 &  \\ 
  Age group & 15-24 years & 705 & 9 & 0 \\ 
   & 25-44 years & 3619 & 48 &  \\ 
   & 45-64 years & 2516 & 33 &  \\ 
   & Minimum 65 years & 745 & 10 &  \\ 
  Purpose of the trip & Vacation, leisure, or recreation & 7272 & 96 & 0 \\ 
   & Some other purpose? & 313 & 4 &  \\ 
  Mode of transportation & Ferry & 2818 & 37 & 0 \\ 
   & Airplane & 4767 & 63 &  \\ 
  Travel group & I travel alone & 1257 & 17 & 109 \\ 
   & I travel with my spouse/significant other & 2547 & 34 &  \\ 
   & I travel with my family, relatives or friends & 3489 & 47 &  \\ 
   & Other & 183 & 2 &  \\ 
  Accommodation & Hotel or hostel & 4689 & 62 & 16 \\ 
   & Rental cottage or apartment (rented privately or from an intermediary, e g  Airbnb or booking com) & 1669 & 22 &  \\ 
   & With friends or relatives (incl  couchsurfing) & 524 & 7 &  \\ 
   & Other & 687 & 9 &  \\ 
  Experienced nature & No & 2954 & 39 & 30 \\ 
   & Yes & 4601 & 61 &  \\ 
  Experienced sports & No & 4941 & 65 & 30 \\ 
   & Yes & 2614 & 35 &  \\ 
  Experienced wellbeing & No & 5164 & 68 & 30 \\ 
   & Yes & 2391 & 32 &  \\ 
  Experienced culture & No & 4518 & 60 & 30 \\ 
   & Yes & 3037 & 40 &  \\ 
  Experienced city life & No & 5153 & 68 & 30 \\ 
   & Yes & 2402 & 32 &  \\ 
  Experienced events & No & 7163 & 95 & 30 \\ 
   & Yes & 392 & 5 &  \\ 
  Experienced shopping & No & 5526 & 73 & 30 \\ 
   & Yes & 2029 & 27 &  \\ 
  Experienced road trip & No & 6321 & 84 & 30 \\ 
   & Yes & 1234 & 16 &  \\ 
  Over 50km trips & Yes & 4126 & 55 & 124 \\ 
   & No & 3335 & 45 &  \\ 
   \hline
\end{tabular}
\endgroup
\label{tab:cat_vars_M0}
\end{table}

\begin{table}[ht!!]
\centering\caption{Work-related trips data ($n$=3376): Frequencies and proportions of categorical variables.} 
\begingroup\tiny
\begin{tabular}{ccccc}
\\
  \hline
Variable & Categories & Frequencies & Proportions (\%) & Missing observations \\ 
  \hline
Main destination & Helsinki & 1742 & 52 & 48 \\ 
   & Helsinki metropolitan area & 324 & 10 &  \\ 
   & Coast and archipelago & 647 & 19 &  \\ 
   & Lakeland & 524 & 16 &  \\ 
   & Lapland & 91 & 3 &  \\ 
  Quarter & 1 & 762 & 23 & 0 \\ 
   & 2 & 1323 & 39 &  \\ 
   & 3 & 618 & 18 &  \\ 
   & 4 & 673 & 20 &  \\ 
  Gender & Male & 2384 & 72 & 71 \\ 
   & Other than male & 921 & 28 &  \\ 
  Age group & 15-24 years & 188 & 6 & 0 \\ 
   & 25-44 years & 1686 & 50 &  \\ 
   & 45-64 years & 1408 & 42 &  \\ 
   & Minimum 65 years & 94 & 3 &  \\ 
  Purpose of the trip & Study & 128 & 4 & 0 \\ 
   & Meeting or work trip in the service of a non-Finnish employer & 969 & 29 &  \\ 
   & Conference or congress or fair & 688 & 20 &  \\ 
   & Work performed in Finland for a Finnish employer & 919 & 27 &  \\ 
   & Some other work-related reason & 672 & 20 &  \\ 
  Mode of transportation & Ferry & 1271 & 38 & 0 \\ 
   & Airplane & 2105 & 62 &  \\ 
  Travel group & I travel alone & 2236 & 68 & 64 \\ 
   & I travel with my spouse/significant other & 176 & 5 &  \\ 
   & I travel with my family, relatives or friends & 235 & 7 &  \\ 
   & Other & 665 & 20 &  \\ 
  Accommodation & Hotel or hostel & 1991 & 59 & 5 \\ 
   & Rental cottage or apartment (rented privately or from an intermediary, e g  Airbnb or booking com) & 404 & 12 &  \\ 
   & Housing provided by an employer & 586 & 17 &  \\ 
   & Other & 390 & 12 &  \\ 
  Experienced nature & No & 2589 & 77 & 9 \\ 
   & Yes & 778 & 23 &  \\ 
  Experienced sports & No & 2824 & 84 & 9 \\ 
   & Yes & 543 & 16 &  \\ 
  Experienced wellbeing & No & 2734 & 81 & 9 \\ 
   & Yes & 633 & 19 &  \\ 
  Experienced culture & No & 2725 & 81 & 9 \\ 
   & Yes & 642 & 19 &  \\ 
  Experienced city life & No & 2524 & 75 & 9 \\ 
   & Yes & 843 & 25 &  \\ 
  Experienced events & No & 3202 & 95 & 9 \\ 
   & Yes & 165 & 5 &  \\ 
  Experienced shopping & No & 2560 & 76 & 9 \\ 
   & Yes & 807 & 24 &  \\ 
  Experienced road trip & No & 3039 & 90 & 9 \\ 
   & Yes & 328 & 10 &  \\ 
  Over 50km trips & Yes & 1644 & 50 & 68 \\ 
   & No & 1664 & 50 &  \\ 
   \hline
\end{tabular}
\endgroup
\label{tab:cat_vars_M1}
\end{table}

\clearpage
\section{Estimates}\label{app_est}


\begin{table}[ht!!]
\centering
\caption{Context group, average causal effects (ACE, in euros) of the extended length of stay on the
total expenditure, and the corresponding 95 \% credible intervals (CI).} 

\begingroup\scriptsize
\begin{tabular}{llrrr}
\\
  \hline
Context group & Length of stay (increase) & ACE & Lower 95 \% CI & Upper 95 \% CI \\ 
  \hline
Personal & 1 (+1) & 142.23 & 123.22 & 161.23 \\ 
  Personal & 2 (+1) & 109.12 & 85.48 & 133.45 \\ 
  Personal & 3 (+1) & 159.37 & 122.28 & 196.94 \\ 
  Personal & 4 (+1) & 110.78 & 58.69 & 164.16 \\ 
  Personal & 5 (+1) & 89.96 & 29.05 & 156.47 \\ 
  Personal & 6 (+1) & 139.94 & 72.76 & 207.76 \\ 
  Personal & 7 (+1) & 100.00 & 24.61 & 190.42 \\ 
  Personal & 8 (+1) & 71.75 & 12.38 & 159.24 \\ 
  Personal & 9 (+1) & 71.50 & 12.16 & 161.30 \\ 
  Personal & 10 (+1) & 54.28 & 6.74 & 135.40 \\ 
  Personal & 11 (+1) & 65.50 & 9.39 & 159.40 \\ 
  Personal & 12 (+1) & 58.90 & 7.83 & 156.59 \\ 
  Personal & 13 (+1) & 130.12 & 28.63 & 270.65 \\ 
  Personal & 14 (+1) & 89.17 & 12.22 & 237.56 \\ 
  Work-related & 1 (+1) & 142.52 & 107.98 & 179.08 \\ 
  Work-related & 2 (+1) & 174.40 & 126.81 & 224.74 \\ 
  Work-related & 3 (+1) & 125.53 & 63.40 & 189.01 \\ 
  Work-related & 4 (+1) & 63.06 & 13.19 & 130.35 \\ 
  Work-related & 5 (+1) & 78.64 & 14.23 & 164.41 \\ 
  Work-related & 6 (+1) & 101.25 & 22.49 & 207.23 \\ 
  Work-related & 7 (+1) & 84.50 & 14.31 & 195.47 \\ 
  Work-related & 8 (+1) & 115.81 & 19.93 & 247.86 \\ 
  Work-related & 9 (+1) & 92.23 & 14.02 & 219.90 \\ 
  Work-related & 10 (+1) & 101.47 & 14.55 & 246.48 \\ 
  Work-related & 11 (+1) & 115.02 & 16.18 & 270.78 \\ 
  Work-related & 12 (+1) & 112.24 & 15.20 & 271.19 \\ 
  Work-related & 13 (+1) & 95.56 & 12.02 & 254.27 \\ 
  Work-related & 14 (+1) & 146.78 & 20.24 & 383.92 \\ 
   \hline
\end{tabular}
\endgroup
\label{tab:ACE}
\end{table}


\begin{table}[ht!!]
\centering\caption{Context group, average causal effects (ACE, in euros) of the extended length of stay on the
total expenditure, and the corresponding 95 \% credible intervals (CI) in the case of omitted variable bias (OVB).} 
\begingroup\scriptsize
\begin{tabular}{llrrr}
\\
  \hline
Context group & Length of stay (increase) & ACE & Lower 95 \% CI & Upper 95 \% CI \\ 
  \hline
Personal (OVB included) & 1 (+1) & 153.47 & 117.02 & 192.66 \\ 
  Personal (OVB included) & 2 (+1) & 120.36 & 78.98 & 161.18 \\ 
  Personal (OVB included) & 3 (+1) & 170.61 & 120.99 & 221.88 \\ 
  Personal (OVB included) & 4 (+1) & 122.02 & 59.18 & 185.88 \\ 
  Personal (OVB included) & 5 (+1) & 101.21 & 32.16 & 172.89 \\ 
  Personal (OVB included) & 6 (+1) & 151.18 & 76.44 & 228.88 \\ 
  Personal (OVB included) & 7 (+1) & 111.24 & 25.18 & 209.50 \\ 
  Personal (OVB included) & 8 (+1) & 82.99 & 13.02 & 175.95 \\ 
  Personal (OVB included) & 9 (+1) & 82.74 & 13.34 & 177.27 \\ 
  Personal (OVB included) & 10 (+1) & 65.52 & 6.41 & 150.71 \\ 
  Personal (OVB included) & 11 (+1) & 76.74 & 8.90 & 174.06 \\ 
  Personal (OVB included) & 12 (+1) & 70.14 & 5.33 & 170.28 \\ 
  Personal (OVB included) & 13 (+1) & 141.36 & 33.90 & 284.08 \\ 
  Personal (OVB included) & 14 (+1) & 100.41 & 12.69 & 249.64 \\ 
  Work-related (OVB included) & 1 (+1) & 144.16 & 107.07 & 183.48 \\ 
  Work-related (OVB included) & 2 (+1) & 176.04 & 126.69 & 226.98 \\ 
  Work-related (OVB included) & 3 (+1) & 127.17 & 62.31 & 191.91 \\ 
  Work-related (OVB included) & 4 (+1) & 64.70 & 12.62 & 131.90 \\ 
  Work-related (OVB included) & 5 (+1) & 80.28 & 14.90 & 166.57 \\ 
  Work-related (OVB included) & 6 (+1) & 102.90 & 22.35 & 209.70 \\ 
  Work-related (OVB included) & 7 (+1) & 86.15 & 14.95 & 196.97 \\ 
  Work-related (OVB included) & 8 (+1) & 117.45 & 21.59 & 250.94 \\ 
  Work-related (OVB included) & 9 (+1) & 93.87 & 14.19 & 221.06 \\ 
  Work-related (OVB included) & 10 (+1) & 103.11 & 14.93 & 250.33 \\ 
  Work-related (OVB included) & 11 (+1) & 116.66 & 16.80 & 274.41 \\ 
  Work-related (OVB included) & 12 (+1) & 113.88 & 15.91 & 273.17 \\ 
  Work-related (OVB included) & 13 (+1) & 97.20 & 11.92 & 257.98 \\ 
  Work-related (OVB included) & 14 (+1) & 148.43 & 21.20 & 387.59 \\ 
   \hline
\end{tabular}
\endgroup
\label{tab:ACE_ovb}
\end{table}

\end{appendices}

\end{document}